\documentclass[aps,prl,twocolumn,groupedaddress,showpacs]{revtex4}
\usepackage{graphicx}
\draft
\begin{document}
\title{Theory of Phonon Hall Effect in Paramagnetic Dielectrics}
\author{L. Sheng$^{1}$, D. N. Sheng$^2$ and C. S. Ting$^1$}
\address{
$^1$Department of Physics and Texas Center for Superconductivity,
University of Houston, Houston, Texas 77204\\
$^2$Department of Physics and Astronomy, California State
University, Northridge, California 91330 }

\begin{abstract}
Based upon spin-lattice interaction, we propose a theoretical
model for the phonon Hall effect in paramagnetic dielectrics. The
thermal Hall conductivity is calculated by using the Kubo formula.
Our theory reproduces the essential experimental features of the
phonon Hall effect discovered recently in ionic dielectric
Tb$_3$Ga$_5$O$_{12}$, including the sign, magnitude and linear
magnetic field dependence of the thermal Hall conductivity.
\end{abstract}

\mbox{}\\

\pacs{66.70.+f, 72.20.Pa, 72.15.Gd, 72.10.Bg} \maketitle


When an electrical current flows through a conductor with
direction perpendicular to an applied magnetic field, a transverse
electrical current may be generated in the third perpendicular
direction. This is well known as the Hall effect, and is due to
the electromagnetic Lorentz force on the charge carriers.
Accompanied transverse heat current also flows in the conductor,
simply because the charge carriers carry energy, known as the
Righi-Leduc effect. Two interesting variants of the conventional
Hall effect is the anomalous Hall effect (AHE) in
ferromagnets~\cite{AHE} and the spin Hall effect in nonmagnetic
conductors~\cite{SHE0,SHE1}, where the electron spin-orbit
coupling plays an essential role. The AHE is characterized by an
anomalous contribution to the Hall resistivity coming solely from
the magnetization. Intuitively, one would not expect a Hall effect
for phonons, which do not carry charges and do not couple to the
magnetic field directly. Remarkably, by applying a magnetic field
perpendicular to a heat current flowing through a sample of the
paramagnetic material Tb$_3$Ga$_5$O$_{12}$, Strohm, Rikken and
Wyder observed very recently a temperature difference of up to 200
microkelvin between the sample edges in the third perpendicular
direction~\cite{PHE}. Since Tb$_3$Ga$_5$O$_{12}$ is a dielectric,
and so the Righi-Leduc effect can be ruled out. The temperature
difference is attributed to the phonon Hall effect
(PHE)~\cite{PHE}, which becomes another intriguing and puzzling
phenomenon in solid state physics.

At the experimental low temperature 5.45K~\cite{PHE}, excitation
of optical phonons is unlikely, and thermal conduction should be
carried by acoustic phonons. While Tb$_3$Ga$_5$O$_{12}$ is an
ionic material, in a perfect lattice, each unit cell must be
charge neutral. In the acoustic phonon modes, each unit cell
vibrates as a rigid object without relative displacements between
its constituting atoms~\cite{SOLID}, and hence does not acquire a
net Lorentz force in a magnetic field. Theoretical understanding
of the physical mechanism underlying the PHE is so far absent.

In this Letter, we propose a theoretical model based upon the
Raman spin-lattice interaction for the PHE. The PHE is discussed
to be a phonon analogue to the AHE. The thermal Hall conductivity
of the phonons in the clean limit is calculated by using the Kubo
formula. The theory can explain the essential features of the
experimental data for Tb$_3$Ga$_5$O$_{12}$, including the sign,
magnitude and linear magnetic field dependence of the thermal Hall
conductivity.

We consider a sample of a paramagnetic dielectric with volume
${\cal V}$, which has a cubic lattice structure with symmetry axes
parallel to the axes of the coordinate system. For simplicity, we
assume that only one rare-earth ion in each unit cell is
paramagnetic. The relevant model Hamiltonian is written as
\begin{equation}
H=\sum\limits_{q,\sigma}\hbar\omega_{q\sigma}a_{q\sigma}^\dagger
a_{q\sigma}+V\ ,\label{HAMIL}
\end{equation}
where $a^\dagger_{q\sigma}$ creates an acoustic phonon of
wavevector ${\bf q}$ and polarization $\sigma$. The well-known
Debye model is used to describe the acoustic phonons. We designate
$\sigma=0$ for the longitudinal acoustic (LA) phonons with
dispersion relation $\omega_{q0}=c_{\mbox{\tiny L}}q$, and
$\sigma=1$ and $2$ for the transverse acoustic (TA) phonons with
$\omega_{q1}=\omega_{q2}=c_{\mbox{\tiny T}}q$. $V$ represents the
interaction between the phonons and the electronic spins and
orbital angular momenta of the paramagnetic
ions~\cite{crystalline,pheno0,pheno1,pheno2,pheno3,sph0}, often
called the spin-lattice interaction. For Tb$_3$Ga$_5$O$_{12}$, the
rare-earth ions Tb$^{3+}$ are paramagnetic with large magnetic
moments and may be responsible for the spin-lattice interaction.
Since microscopic calculation of the spin-lattice interaction is
difficult, phenomenological description based upon symmetry
considerations is usually
employed~\cite{pheno0,pheno1,pheno2,pheno3,sph0}. Two most
important spin-lattice interaction processes have been extensively
studied in the
past~\cite{crystalline,pheno0,pheno1,pheno2,pheno3,sph0}, i.e.,
the modulation interaction and the Raman interaction. The
modulation interaction stands for the direct modulation of the
inter-atomic spin-spin interaction due to lattice vibrations. The
Raman interaction represents the second-order transition between
the Kramers degenerate ground states of an ion through the
intermediary of the excited states caused by the time-dependent
variation of the crystalline field. We note that, for
Tb$_3$Ga$_5$O$_{12}$, magnetic ordered states do not occur down to
very low temperature 0.2$K$~\cite{GARNETs}, suggesting that the
inter-atomic spin-spin interaction is possibly very weak. We will
neglect the spin-spin interaction, and focus on the Raman
interaction. The Raman interaction is known to dominate the
spin-lattice relaxation in many ionic
insulators~\cite{crystalline,pheno0,pheno1,pheno2}.

The electron spin-orbit coupling of a rare-earth ion is usually
stronger than the crystalline field. By considering the spin-orbit
coupling and intra-atomic Coulomb interaction first, the spins and
orbital angular momenta of the outer-shell electrons shall form a
total angular momentum ${\bf J}$. The ground states are a ${\bf
J}$ multiplet, which further split in the presence of the
crystalline field~\cite{pheno0,pheno1}. In order to develop a
transparent theory, we study a relatively simple case that was
often considered in
literatures~\cite{pheno0,pheno1,pheno2,pheno3,sph0}. We assume
that all the ground-state degeneracies of the ion except for the
Kramers one are lifted by the crystalline field in such a manner
that the energy difference $E_{1}$ between the lowest excited
states and the ground states is greater than the Debye energy
$\hbar\omega_{\mbox{\tiny D}}$. In this case, one can obtain for
the Raman interaction~\cite{pheno3,sph0} $V= K_{0}\sum_{m}{\bf
s}_m\cdot\mbox{\boldmath{$\Omega$}}_m$, where $K_{0}$ is a
positive coupling constant, the 1/2 isospin ${\bf s}_m$ describes
the ground-state Kramers doublet, and
$\mbox{\boldmath{$\Omega$}}_m={\bf u}_m\times{\bf p}_m$ is the
center-of-mass angular momentum of the unit cell with ${\bf u}_m$
and ${\bf p}_m$ the center-of-mass displacement and momentum. The
Raman interaction is taken to be isotropic by virtue of the cubic
lattice symmetry. At low temperatures, where thermal excitation of
the ion into the excited states is virtually impossible, the
primary contribution to the magnetization comes from the
ground-state doublet. The Kramers doublet carry opposite magnetic
moments, which split in the presence of a magnetic field ${\bf B}$
and give rise to a magnetization ${\bf M}$. It is easy to prove
that the ensemble average of the isospin is proportional to the
magnetization, namely, $\langle{\bf s}_m\rangle=c{\bf M}$ with $c$
the proportionality coefficient~\cite{note1}. Under the mean-field
approximation, the Raman interaction reduces to
\begin{equation}
V=K\sum_{m}{\bf M}\cdot\mbox{\boldmath{$\Omega$}}_m\
,\label{RAMAN_APPRX}
\end{equation}
where $K=cK_{0}$, and $KM$ has the units of frequency. While Eq.\
(\ref{RAMAN_APPRX}) is obtained in the relatively simple case, it
represents a minimal form of possible interaction between
magnetization and phonons, and may also serve as a reasonable
hypothesis for a phenomenological theory of the PHE in general
paramagnetic dielectrics, similar to the theory of the
AHE~\cite{AHE}.

\begin{figure}
\includegraphics[width=1.9in]{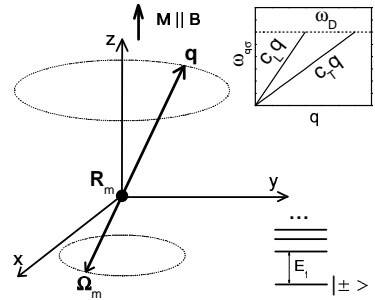}
\caption{Lowest-energy configuration between the phonon wave
vector ${\bf q}$ and the center-of-mass angular momentum
$\mbox{\boldmath{$\Omega$}}_m$ of a unit cell. Here, ${\bf R}_m$
represents the equilibrium position of the unit cell, which is
taken to be at the origin. Top right panel shows the dispersion
relations of the LA and TA phonons in the Debye model. Inset at
the bottom right is a hypothetical level graph of a paramagnetic
ion with $\vert\pm\rangle$ as the ground-state doublet.}
\end{figure}
In order to illustrate the basic mechanism of the PHE, we consider
the vibration of the $m$-th unit cell in the phonon modes of a
given wave vector ${\bf q}$, as shown in Fig.\ 1. The sound speed
$c_{\mbox{\tiny T}}$ of the TA phonons is generally smaller than
that of the LA phonons $c_{\mbox{\tiny L}}$~\cite{HANDBOOK},
meaning that the vibration of the unit cell in the transverse
directions is lower in energy than that in the longitudinal
direction. This yields a correlation between the angular momentum
$\mbox{\boldmath{$\Omega$}}_m$ and phonon momentum $\hbar{\bf q}$
that parallel and antiparallel alignments between
$\mbox{\boldmath{$\Omega$}}_m$ and $\hbar{\bf q}$ are
energetically favorable. We may regard
$\mbox{\boldmath{$\Omega$}}_m$ as an internal ``spin'' degree of
freedom of the phonons. Then the correlation between
$\mbox{\boldmath{$\Omega$}}_{m}$ and $\hbar{\bf q}$ plays the role
of a ``spin-orbit'' coupling, effectively similar to the Luttinger
spin-orbit coupling~\cite{LUTT}. Besides, the Raman interaction
Eq.\ (\ref{RAMAN_APPRX}) is a counterpart to the exchange coupling
between carrier spins and magnetization in the AHE
systems~\cite{AHE}. With these two essential ingredients, we can
expect the PHE to occur in the present system, as a bosonic
analogue to the AHE~\cite{AHE}. In Fig.\ 1, the lowest-energy
configuration between $\mbox{\boldmath{$\Omega$}}_{m}$ and
$\hbar{\bf q}$ is shown. It has been considered that, when ${\bf
M}\neq 0$ and the Raman interaction arises as a weak perturbation,
the degenerate TA modes split into two circularly polarized modes
with nonvanishing $\mbox{\boldmath{$\Omega$}}_m$ parallel and
antiparallel to $\hbar{\bf q}$, respectively, and the one with
$\Omega_{m}^{z}<0$ is lower in energy. The splitting occurs in
this way because $\vert\mbox{\boldmath{$\Omega$}}_m\vert$ is
maximized in the circularly polarized modes, which is favorable
for minimizing the Raman interaction.

We start the rigorous calculation by rewriting the Raman
interaction Eq.\ (\ref{RAMAN_APPRX}) in the second-quantization
representation
\begin{equation}
V=\frac{1}{2}\sum\limits_{{\bf
q},\sigma,\sigma'}\Delta_{q\sigma\sigma'}\sqrt{\frac{\omega_{q\sigma'}}{\omega_{q\sigma}}}
(a_{-q\sigma}+a_{q\sigma}^\dagger)
(a_{q\sigma'}-a_{-q\sigma'}^\dagger)\ ,\label{INTERACTION}
\end{equation}
where $\Delta_{q\sigma\sigma'}=-i\hbar K{\bf M}\cdot(\hat{\bf
e}^*_{q\sigma}\times \hat{\bf e}_{q\sigma'})$ with $\hat{\bf
e}_{q\sigma}$ the phonon polarization vector, and the phase
convention ${\bf e}^{*}_{-q\sigma}={\bf e}_{q\sigma}$ is adopted.
The thermal Hall conductivity $\kappa_{xy}$ can be calculated by
using the Kubo formula~\cite{Kubo}
\begin{eqnarray}
\kappa_{xy}&=&\frac{{\cal V}}{T}\int_0^{{\hbar}/{k_{\mbox{\tiny
B}}T}}d\lambda\int_0^\infty dt\langle J_{\mbox{\tiny
E}}^{x}(-i\lambda)J_{\mbox{\tiny E}}^{y}(t)\rangle\ ,\label{KUBO0}
\end{eqnarray}
where $J_{\mbox{\tiny E}}^{x}$ is the $x-$component of the energy
flux operator ${\bf J}_{\mbox{\tiny E}}$ of the phonons, and ${\bf
J}_{\mbox{\tiny E}}(t)=e^{iHt/\hbar}{\bf J}_{\mbox{\tiny
E}}e^{-iHt/\hbar}$. The complete expression for ${\bf
J}_{\mbox{\tiny E}}$ for free phonons was derived by
Hardy~\cite{EFlux}, which consists of some harmonic terms and
cubic terms. In the same essence as the harmonic approximation to
the real lattice Hamiltonian in the fundamental phonon
theory~\cite{SOLID}, it is sufficient to retain the harmonic terms
at low temperatures. We extend the derivation~\cite{EFlux} to include
the contribution from the Raman interaction
\begin{eqnarray}
{\bf J}_{\mbox{\tiny E}}=\frac{1}{2{\cal
V}}\sum_{mn\alpha\beta}({\bf R}_m-{\bf
R}_n)\Phi^{\alpha\beta}({\bf R}_m-{\bf
R}_n)u_{m}^{\alpha}v_{n}^{\beta}\ ,
\end{eqnarray}
where $u_{m}^{\alpha}$ and $v_{m}^{\alpha}$ with $\alpha=x$, $y$
and $z$ are the $\alpha$-components of the center-of-mass
displacement ${\bf u}_m$ and velocity ${\bf v}_m$ of the $m$-th
unit cell, respectively, and $\Phi^{\alpha\beta}({\bf R}_m-{\bf
R}_n)$ are the stiffness matrix elements of the lattice with ${\bf
R}_{m}$ the equilibrium position of the unit cell. The
center-of-mass velocity is ${\bf v}_{m}={\bf p}_m/M_{c}+K({\bf
M}\times{\bf u}_m)$ with $M_{c}$ the mass of a unit cell, where
the second term originates from the Raman interaction. By using
the basic relation~\cite{SOLID} $\Phi^{\alpha\beta}({\bf
q})=M_{c}\sum_{\sigma}
(\omega_{q\sigma}^2\hat{e}_{q\sigma}^{\alpha}\hat{e}_{q\sigma}^{*\beta})$,
the energy flux is derived to be ${\bf J}_{\mbox{\tiny E}}={\bf
J}_{\mbox{\tiny E}}^{(0)}+{\bf J}_{\mbox{\tiny E}}^{(1)}$, where
\begin{eqnarray}
{\bf J}_{\mbox{\tiny E}}^{(0)}&=&\frac{1}{2{\cal
V}}\sum\limits_{{\bf q},\sigma,\sigma'}{\bf
j}_{q\sigma\sigma'}\sqrt{\frac{\omega_{q\sigma'}}{\omega_{q\sigma}}}(a_{-q\sigma}+
a_{q\sigma}^{\dagger})\nonumber\\
&\times&(a_{q\sigma'}-a_{-q\sigma'}^{\dagger})\ ,\label{J0}
\end{eqnarray}
\begin{eqnarray}
{\bf J}_{\mbox{\tiny E}}^{(1)}&=&\frac{1}{2{\cal
V}}\sum\limits_{{\bf q},\sigma,\sigma',\sigma''}{\bf
j}_{q\sigma\sigma''}\left(\frac{\Delta_{q\sigma''\sigma'}}
{\hbar\sqrt{\omega_{q\sigma}\omega_{q\sigma'}}}\right)\nonumber\\
&\times&(a_{-q\sigma}+ a_{q\sigma}^{\dagger})
(a_{q\sigma'}+a_{-q\sigma'}^\dagger)\ ,\label{J1}
\end{eqnarray}
with
\begin{eqnarray}
{\bf
j}_{q\sigma\sigma'}&=&\hbar\omega_{q\sigma}\delta_{\sigma\sigma'}\nabla_{q}\omega_{q\sigma}
+\frac{\hbar}{4}( \omega_{q\sigma}^2-\omega_{q\sigma'}^2)\nonumber\\
&\times&\sum\limits_\alpha\left[(\nabla_q\hat{e}_{q\sigma}^{*\alpha})
\hat{e}_{q\sigma'}^{\alpha}-\hat{e}_{q\sigma}^{*\alpha}(\nabla_q
\hat{e}_{q\sigma'}^{\alpha})\right]\ .\label{VEL}
\end{eqnarray}
Here, ${\bf J}_{\mbox{\tiny E}}^{(1)}$ comes from the Raman
interaction.

The energy scale $\hbar KM$ of the Raman interaction is generally
much smaller than the Debye energy $\hbar\omega_{\mbox{\tiny D}}$
as well as the average energy difference between the LA and TA
phonon branches, so that a perturbation treatment of Eq.\
(\ref{INTERACTION}) suffices. Since the two TA branches are
degenerate, according to the degenerate perturbation theory, we
need to find suitable phonon polarization vectors, for which the
direct coupling between the two TA branches
$\Delta_{q12}=\Delta^{*}_{q21}$ vanishes. Such polarization
vectors are obtained as $\hat{\bf e}_{q0}=i\hat{\bf q}$, $\hat{\bf
e}_{q1}=(\hat{\mbox{\boldmath{$\theta$}}}_q+i\hat{\mbox{\boldmath{$\varphi$}}}_q)/\sqrt{2}$
and $\hat{\bf
e}_{q2}=(\hat{\mbox{\boldmath{$\theta$}}}_q-i\hat{\mbox{\boldmath{$\varphi$}}}_q)/\sqrt{2}$,
where $\hat{\bf q}$, $\hat{\mbox{\boldmath{$\theta$}}}_q$ and
$\hat{\mbox{\boldmath{$\varphi$}}}_q$ are the unit vectors
associated with wavevector ${\bf q}=(q,\theta_q,\varphi_q)$ in the
spherical polar coordinate system. We notice that $\hat{\bf
e}_{q1}$ ($\hat{\bf e}_{q2}$) for the TA phonons is a
superposition of two orthogonal linear polarization vectors
$\hat{\mbox{\boldmath{$\theta$}}}_q$ and
$\hat{\mbox{\boldmath{$\varphi$}}}_q$ with a fixed phase shift
$90^{\circ}$ ($-90^{\circ}$) between them. This indicates that the
TA phonons indeed split into two circularly polarized branches
upon the perturbation of the spin-lattice interaction. The
standard nondegenerate perturbation theory is then applied to
calculate the many-particle eigenstates of the system to the
linear order in the Raman interaction. By substitution of the
eigenstates into Eq.\ (\ref{KUBO0}), we consequently obtain for
the thermal Hall conductivity
\begin{equation}
{\kappa}_{xy}=\frac{\gamma k_{\mbox{\tiny
B}}KM}{2\pi^2\overline{c}_s}\left(\frac{k_{\mbox{\tiny
B}}T}{\hbar}\right)\int_0^{\Theta{\mbox{\tiny
D}}/T}\frac{x}{e^{x}-1}dx\ ,\label{HALL_THERMAL}
\end{equation}
where
$\gamma=(5-\delta)(1+\delta)^4/[4\delta^{2}(9+18\delta^3)^{1/3}]$
with $\delta=c_{\mbox{\tiny L}}/c_{\mbox{\tiny T}}$,
$\overline{c}_s$ is the average sound speed defined by
$3/\overline{c}_s^{3}=(1/c_{\mbox{\tiny L}}^{3}+2/c_{\mbox{\tiny
T}}^{3})$ and $\Theta_{\mbox{\tiny D}}=\hbar \omega_{\mbox{\tiny
D}}/k_{\mbox{\tiny
B}}=(6\pi^2/\nu_{0})^{1/3}\hbar\overline{c}_{s}/k_{\mbox{\tiny
B}}$ is the Debye temperature with $\nu_{0}$ the volume of a unit
cell. The next-order correction to $\kappa_{xy}$ will be of the
order of $(KM/\omega_{\mbox{\tiny D}})^{3}$, which can be expected
to be extremely small.

\begin{figure}
\includegraphics[width=2.5in]{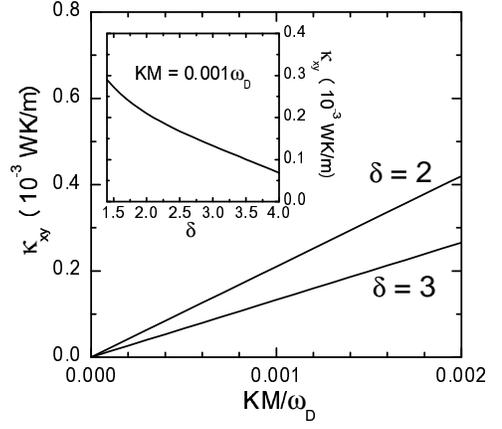}
\caption{Thermal Hall conductivity $\kappa_{xy}$ as a function of
$KM/\omega_{\mbox{\tiny D}}$ for two different values of
$\delta=c_{\mbox{\tiny L}}/c_{\mbox{\tiny T}}$. Inset:
$\kappa_{xy}$ as a function of $\delta$ for
$KM=0.001\omega_{\mbox{\tiny D}}$. Here, $T=5.45$K,
$\hbar\omega_{\mbox{\tiny D}}=0.05$eV and $\nu_{0}=(12$\AA$)^3$.}
\end{figure}
In Fig.\ 2, the calculated $\kappa_{xy}$ at $T=5.45$K is plotted
as a function of $KM/\omega_{\mbox{\tiny D}}$ for two different
values of $\delta$. Here, we set $\hbar\omega_{\mbox{\tiny
D}}=0.05$eV and $\nu_{0}=(12$\AA$)^3$, which correspond to
$\Theta_{\mbox{\tiny D}}\simeq 580$K and $\overline{c}_{s}\simeq
850$m/s. These are either known values for Tb$_3$Ga$_5$O$_{12}$ or
typical values for rare-earth garnets with similar
structures~\cite{GARNETs}. It is clear that the thermal Hall
conductivity $\kappa_{xy}$ is linear in the magnetization $M$.
When the applied magnetic field $B$ is relatively weak, $M$ varies
linearly with $B$, and so does $\kappa_{xy}$. We can expect that,
while $\kappa_{xy}$ remains linear as a function of $M$, it may
possibly become nonlinear as a function of $B$ at very strong
magnetic field. It is worthwhile to test this prediction in
experiment. In the inset of Fig.\ 2, $\kappa_{xy}$ as a function
of $\delta$ is plotted. $\kappa_{xy}$ decreases with increasing
$\delta$. From Eq.\ (\ref{HALL_THERMAL}), we see that
$\kappa_{xy}$ will change sign at $\delta=5$. We notice that Eqs.\
(\ref{J0}) and (\ref{J1}) make comparable opposite contributions
to $\kappa_{xy}$. Their competition accounts for the nontrivial
dependence of $\kappa_{xy}$ on $\delta$. In most materials, the
typical values of $\delta$ are around $2$~\cite{HANDBOOK}, where
$\kappa_{xy}$ is always positive.

\begin{figure}
\includegraphics[width=2.5in]{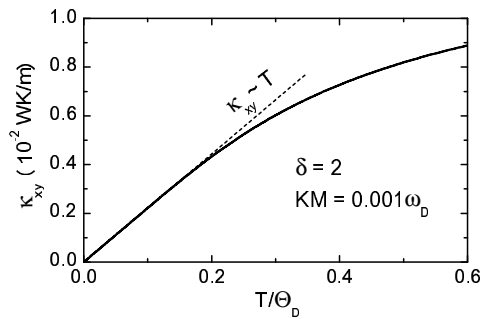}
\caption{$\kappa_{xy}$ as a function of normalized temperature for
$KM=0.001\omega_{\mbox{\tiny D}}$ and $\delta=2$ (solid line).
Other parameters are chosen to be the same as in Fig.\ 2. The
dashed line is a linear fit to $\kappa_{xy}$ at low temperatures.
}
\end{figure}
In Fig.\ 3, $k_{xy}$ for a fixed value of $KM$ is shown as a
function of normalized temperature $T/\Theta_{\mbox{\tiny D}}$. At
low temperatures $T\ll \Theta_{\mbox{\tiny D}}$, we can obtain
from Eq.\ (\ref{HALL_THERMAL}) ${\kappa}_{xy}\simeq (1.64\gamma
k_{\mbox{\tiny B}}KM/2\pi^2\overline{c}_s)(k_{\mbox{\tiny
B}}T/\hbar)$, where the numerical identity $\sum_{n\ge
1}n^{-2}\simeq 1.64$ has been used. From Fig.\ 3, we see that the
above linear dependence of $\kappa_{xy}$ on temperature is well
valid for $T\lesssim 0.2\Theta_{\mbox{\tiny D}}$. From
experimental point of view, it is relatively convenient to
consider fixed magnetic field instead of fixed magnetization. For
paramagnetic materials and relatively weak magnetic field, we may
use the Curie's law $M\propto B/T$. As a result, at fixed magnetic
field, $\kappa_{xy}$ approaches a constant at low temperatures. At
relatively high temperatures $T\gtrsim 0.2\Theta_{\mbox{\tiny
D}}$, it decreases with increasing temperature.

We can show that the present theory reproduces the essential
characteristics of the experimental data~\cite{PHE}. In the
experiment~\cite{PHE}, a constant longitudinal heat current is
driven through the sample, and the temperature difference $\Delta
T$ in the transverse direction across the sample is measured.
Firstly, the present theory predicts $\kappa_{xy}>0$ for typical
values of $\delta$, meaning that, if the magnetic field is along
the $z$-axis and a heat current is driven along the $x$-axis, the
Hall heat current will flow along the $y$-axis, which is
consistent with the experimental relation among the sign of
$\Delta T$ and the directions of the magnetic field and the
driving heat current~\cite{PHE}. Secondly, $\Delta T$ is observed
to be linear in the magnetic field $B$~\cite{PHE}. By analogy with
the charge Hall effect, $\Delta T$ is proportional to the thermal
Hall resistivity
$r_{xy}=\kappa_{xy}/(\kappa_{xx}^2+\kappa_{xy}^2)\simeq
\kappa_{xy}/\kappa_{xx}^2$, where the longitudinal thermal
conductivity $\kappa_{xx}$ depends weakly on $B$~\cite{PHE}. It
follows that the experimental thermal Hall conductivity
$\kappa_{xy}\propto B$. In the present theory, $\kappa_{xy}\propto
B$ at relatively weak magnetic field, which agrees with the
experimental observation. Thirdly, if $KM$ is taken to be about $
10^{-4}\omega_{\mbox{\tiny D}}$, we see from Fig.\ 2 that
$\kappa_{xy}$ is $10^{-5}-10^{-4}$WK/m, which will be comparable
to the experimental value $4.5\times 10^{-5}$WK/m at $B=1$T and
$T=5.45$K. Here, the experimental value of $\kappa_{xy}$ is
deduced from the reported thermal conductivity
$\kappa_{xx}=4.5\times 10^{-1}$WK/m and Hall angle
$\kappa_{xy}/\kappa_{xx}=1\times 10^{-4}\mbox{rad}\mbox{
}$T$^{-1}$~\cite{PHE}. To estimate the Raman coupling strength, we
can use the relation $KM=K_{0}\vert\langle{\bf s}_m\rangle\vert$.
For the experimental material, while the average of the isospin is
affected by many factors, such as the actual crystalline field and
multiple paramagnetic ions in a unit cell, we may conservatively
assume that the effective $\vert\langle{\bf s}_m\rangle\vert$ is
of the order of $0.1\hbar$ at $B=1$T and $T=5.45$K, where the
Zeeman energy is already comparable to $k_{\mbox{\tiny B}}T$. Thus
$K_{0}$ is estimated about $ K_{0}\simeq
10^{-3}\omega_{\mbox{\tiny D}}/\hbar$ or $\hbar^{2}K_{0}\simeq
1$cm$^{-1}$, which can be found to be within the possible range of
the Raman coupling strength in ionic
insulators~\cite{pheno0,pheno1}.

In summary, we have developed a simple theory based upon the Raman
spin-lattice interaction for the PHE, which has been shown in good
agreement with the experimental data. Further investigation by
including disorder effect is highly desirable. Since the PHE is
similar to the AHE, we expect that the thermal Hall conductivity
might be insensitive to weak disorder scattering~\cite{AHE}.

\textbf{Acknowledgment:} This work is supported by ACS-PRF
41752-AC10, Research Corporation Fund CC5643, the NSF
grant/DMR-0307170 (DNS), and a grant from the Robert A. Welch
Foundation under the grant no. E-1146 (CST).


\begin{references}
\bibitem{note1}
By assumption, $E_{1}$ should be much greater than the Zeeman
energy, and so the Zeeman energy admixtures of excited states into
the ground-state doublet are negligible. The $z$-axis is chosen as
the quantization axis for ${\bf J}$ and ${\bf s}_m$. $J^{z}$
projected into the subspace spanned by the doublet can be written
in the general form $J^{z}=\xi_{0}+\xi_{1}({\bf
s}_{m}\cdot\hat{\bf e})$, where $\xi_{0}$ and $\xi_{1}$ are two
constants. Due to time-reversal symmetry, $\xi_{0}$ actually
vanishes. By considering cubic lattice symmetry, the unit vector
$\hat{\bf e}$ is found along the $z$-axis, and hence
$J^{z}=\xi_{1} s_{m}^{z}$. Thus $M^z=\langle g_{\mbox{\tiny
J}}\mu_{\mbox{\tiny B}}J^{z}\rangle/\nu_{0}=g_{\mbox{\tiny
J}}\mu_{\mbox{\tiny B}}\xi_{1}\langle s_{m}^{z}\rangle/\nu_{0}$
with $\nu_{0}$ the volume of a unit cell. Since ${\bf M}$ and
$\langle{\bf s}_{m}\rangle$ are both parallel to the $z$-axis by
symmetry, we can also write $\langle{\bf s}_{m}\rangle=c{\bf M}$
with $c=\nu_0/g_{\mbox{\tiny J}}\mu_{\mbox{\tiny B}}\xi_{1}$.
\end{references}
\end{document}